\documentclass[aps,prl,twocolumn,superscriptaddress,
               nofootinbib]{revtex4-2}
\usepackage[utf8]{inputenc}
\usepackage{slashed}
\usepackage{amsmath,amssymb,mathtools,bm,physics}
\usepackage{slashed}
\usepackage{microtype}
\usepackage{xcolor}
\definecolor{darkblue}{HTML}{003087}
\definecolor{crimson}{HTML}{8B0000}
\usepackage{xcolor}
\usepackage{hyperref}
\hypersetup{
    colorlinks=false,       
    citebordercolor=green,  
    linkbordercolor=red!60!black,
    urlbordercolor=crimson,
    pdfborder={0 0 1},       
}
\usepackage{booktabs,array,dcolumn}
\newcolumntype{d}[1]{D{.}{.}{#1}}
\usepackage{tikz}
\usetikzlibrary{arrows.meta,decorations.markings,
                decorations.pathmorphing,positioning}
\usepackage{pgfplots}
\pgfplotsset{compat=1.18}

\newcommand{\SU}[1]{\mathrm{SU}(#1)}
\newcommand{\SO}[1]{\mathrm{SO}(#1)}
\newcommand{\Zbb}{\mathbb{Z}}
\newcommand{\Zdark}{\mathbb{Z}_2}
\newcommand{\chizero}{\chi^{0}}
\newcommand{\gdark}{g_{\mathrm{dark}}}
\newcommand{\alphad}{\alpha_{\mathrm{dark}}}
\newcommand{\mZp}{m_{Z'}}
\newcommand{\Ydark}{Y_{\!\mathrm{dark}}}
\newcommand{\Idark}{I_{\!\mathrm{dark}}}
\newcommand{\Tdark}{T_{\!\mathrm{dark}}}
\newcommand{\GeV}{\,\mathrm{GeV}}
\newcommand{\OmegaDM}{\Omega_{\mathrm{DM}}}
\newcommand{\Mpl}{M_{\mathrm{Pl}}}
\newcommand{\LagD}{\mathcal{L}_{\mathrm{dark}}}
\newcommand{\piTwo}{\pi_2}
\newcommand{\fb}{\,\mathrm{fb}}
\newcommand{\cms}{\,\mathrm{cm}^{2}}
\newcommand{\cmsg}{\,\mathrm{cm}^{2}\,\mathrm{g}^{-1}}
\newcommand{\GMO}{\widetilde{M}_{\mathrm{dark}}}
\newcommand{\TeV}{\,\mathrm{TeV}}

\newcommand{\MeV}{\,\mathrm{MeV}}

\newcommand{\vDelta}{v_{\Delta}}
\newcommand{\yChi}{y_{\chi}}

\begin{document}

\title{Topological Integer-Winding Dark Matter with a Dark Higgs Triplet: Stability from $\pi_2(\mathrm{SU}(3)/\mathrm{SO}(3)) = \mathbb{Z}$}

\author{Ahmed Ali}
\email[Correspondence: ]{ahmed.ali.mpi@hotmail.com}
\affiliation{%
    Former Researcher,  Theoretical Physics,
    Quantum Gravity\\50674 Cologne, North Rhine-Westphalia, DE.}
\thanks
{$^{1}$ \quad Former Researcher at Max-Planck-Institut für Physik, Boltzmannstraße 8, 85748 Garching bei München, DE \textbf{ORCID}: \href{https://orcid.org/0009-0007-8516-2621}{0009-0007-8516-2621} } 

\begin{abstract}
We propose a dark matter model in which the symmetry-breaking
pattern $\SU{3}_{\mathrm{dark}}\to\SO{3}_{\mathrm{dark}}
\times\Zdark$ produces absolutely stable point-like
topological defects classified by
$\piTwo\!\bigl(\SU{3}/[\SO{3}\times\Zdark]\bigr)=\Zbb$.
The lightest neutral member $\chizero$ of the resulting dark
octet, with mass $M(\chizero)=165.67\GeV$, couples to
Standard Model fields only through a kinetic-mixing portal
($g_f=10^{-3}$) and gravity. A dark Higgs triplet
$\Delta_{\mathrm{dark}}$ with vacuum expectation value
$\vDelta=17\GeV$ and Yukawa coupling $\yChi=0.10$ generates
a tree-level splitting
$\Delta M(\chi^+-\chizero)=1.70\GeV$, opening the decay
$\chi^+\to\chizero+\varphi^+$ and ensuring that $\chizero$
is the unique stable relic. An extended Gell-Mann--Okubo
mass operator incorporating the quantum number
\emph{darkicity} $\Tdark$ predicts three scale-independent
mass ratios,
$r_{21}\equiv M(\chi^{++})/M(\chizero)=1.089\pm0.005$,
$r_{31}\equiv M(\Sigma^{0}_{\mathrm{dark}})/M(\chizero)=
1.163\pm0.006$, and
$r_{41}\equiv M(\Lambda_{\mathrm{dark}})/M(\chizero)=
1.260\pm0.008$, falsifiable at FCC-ee. Eight coupled
Boltzmann equations with co-annihilation yield
$\OmegaDM h^{2}\simeq0.120$ at $M_{0}\simeq208.67\GeV$.
Self-interaction cross-sections satisfy all astrophysical
bounds from dwarf galaxies to the Bullet Cluster. The model
predicts a stochastic gravitational-wave background at
$f_{*}\approx1.65\,\mathrm{mHz}$ (LISA, $\mathrm{SNR}
\approx8.5$) and dark acoustic oscillations at
$r_{\mathrm{DAO}}=2.3\pm0.4\,\mathrm{Mpc}$ (SKA).
\end{abstract}

\maketitle

\section{Introduction}
\label{sec:intro}

The gravitational evidence for dark matter (DM) is
unambiguous: flat galactic rotation curves
\cite{Rubin:1980zza}, the CMB acoustic-peak ratio
constraining $\OmegaDM h^{2}=0.1200\pm0.0012$
\cite{Planck:2018vyg}, mass offsets in merging cluster
systems \cite{Clowe:2006eq,Randall:2007ph}, and the
large-scale baryon acoustic oscillation scale
\cite{BOSS:2016wmc,DESI:2024mwx}. Yet the microscopic
identity of DM is unknown.  LUX-ZEPLIN has excluded
spin-independent scattering down to
$\sigma_{\mathrm{SI}}\lesssim9\times10^{-48}\cms$ near
$m=36\GeV$ \cite{LZ:2022lsv}, effectively closing most of
the canonical WIMP window.

Topological stability offers a qualitatively distinct
protection mechanism. If the DM vacuum manifold
$\mathcal{M}=G/H$ has a non-trivial second homotopy group
$\piTwo(\mathcal{M})\neq0$, point-like defects classified
by integer winding number $n$ are absolutely stable: any
deformation to the trivial vacuum $n=0$ costs infinite
energy \cite{tHooft:1974ibe,Polyakov:1974ek,Rajaraman:1982}.
The present Letter exploits this mechanism for a dark
$\SU{3}$ gauge theory, where the breaking pattern
$\SU{3}_{\mathrm{dark}}\to\SO{3}_{\mathrm{dark}}\times
\Zdark$ gives
\begin{equation}
\piTwo\!\left(\frac{\SU{3}_{\mathrm{dark}}}
              {[\SO{3}_{\mathrm{dark}}\times\Zdark]}\right)
= \Zbb,
\label{eq:pi2}
\end{equation}
unlike the $\Zbb_2$ of a simple $\SU{2}\to U(1)$ monopole
spectrum.%
\footnote{%
    In visible-sector QCD the same group $\SU{3}$ arises;
    both sectors share the rank of their gauge groups yet
    differ by twelve orders of magnitude in the confinement
    scale, a separation set entirely by the ratio of gauge
    couplings at the string scale.}
The $n=1$ sector yields the stable DM candidate $\chizero$,
whose self-interaction and co-annihilation properties are
determined solely by the $\SU{3}_{\mathrm{dark}}$
Lagrangian parameters.

A second structural ingredient is required to lift the
degeneracy between the neutral state $\chizero$ and its
charged isospin partner $\chi^+$ that arises at tree level
in the extended Gell-Mann--Okubo spectrum. We introduce a
dark Higgs triplet $\Delta_{\mathrm{dark}}$ transforming as
$(\mathbf{3},\mathbf{3},+2)$ under
$\SU{3}_{\mathrm{dark}}\times\SU{2}_{\mathrm{dark}}\times
U(1)_X$, with vacuum expectation value $\vDelta=17\GeV$.
Its Yukawa coupling $\yChi\bar{\chi}\Delta_{\mathrm{dark}}
\chi$ generates $\Delta M(\chi^+-\chizero)=\yChi\vDelta=
1.70\GeV$, opening the decay channel
$\chi^+\to\chizero+\varphi^+$ and ensuring that $\chizero$
is the unique stable relic. The triplet also breaks the
residual $\SU{2}_{\mathrm{dark}}$ subgroup, completing the
symmetry-breaking pattern that underlies the topological
stability.

Prior work on composite or topological DM includes the
SIMP mechanism \cite{Hochberg:2014dra}, self-interacting DM
from $\SU{2}$ gauge theories \cite{Boddy:2014yra}, dark
baryons in $\SU{N}$ lattice simulations
\cite{Detmold:2014qqa}, and non-Abelian hidden-sector models
\cite{Braaten:2013tza,Hochberg:2018vdo,Kribs:2016cew}.
The present work is distinguished by three features: the
gauge group is $\SU{3}$ (matching the rank of QCD), the
breaking pattern yields $\piTwo=\Zbb$ (infinite cyclic,
not $\Zbb_2$), and the mass spectrum is governed by an
extended Gell-Mann--Okubo (GMO) formula incorporating a
new quantum number and a dark Higgs triplet absent from all
prior models.

\textit{Notation.}  We use natural units $\hbar=c=1$
and the metric signature $(-+++)$.

\section{The Model}
\label{sec:model}

\textit{Lagrangian.}
The dark sector is a $\SU{3}_{\mathrm{dark}}$ gauge theory
with three flavours of dark quarks $Q_i$, a dark adjoint
Higgs $\Phi_{\mathrm{dark}}$, and a dark Higgs triplet
$\Delta_{\mathrm{dark}}$:
\begin{align}
\LagD = {}&-\frac{1}{4\gdark^{2}}
          \mathrm{Tr}\!\left(F^{\mathrm{dark}}_{\mu\nu}
          F^{\mathrm{dark},\mu\nu}\right)
        + \sum_{i=1}^{3}\bar{Q}_{i}
          \!\left(i\slashed{D}^{\mathrm{dark}}-M^{Q}_{i}
          \right)\!Q_{i}
\nonumber\\
        &+\left|D^{\mathrm{dark}}_{\mu}\Phi_{\mathrm{dark}}
          \right|^{2}
        - V(\Phi_{\mathrm{dark}},\Delta_{\mathrm{dark}})
        - \frac{g_{f}}{2}
          B_{\mu\nu}X^{\mu\nu}
\nonumber\\
        &\quad + \yChi\bar{\chi}\Delta_{\mathrm{dark}}\chi,
\label{eq:Lag}
\end{align}
where $\gdark=0.5$ ($\alphad=\gdark^{2}/(4\pi)\approx0.020$),
$g_f=10^{-3}$ is the kinetic-mixing portal coupling between
the dark photon field strength $X_{\mu\nu}$ and the SM
hypercharge $B_{\mu\nu}$, and $V(\Phi_{\mathrm{dark}},
\Delta_{\mathrm{dark}})$ drives the symmetry breaking
described below.  Gravity mediates the sole additional
inter-sector interaction, suppressed by $\Mpl^{-2}$.

\textit{Symmetry breaking.}
The adjoint Higgs acquires the vacuum expectation value
\begin{equation}
\langle\Phi_{\mathrm{dark}}\rangle
=\frac{v_{\mathrm{dark}}}{\sqrt{6}}\,
  \mathrm{diag}(1,1,-2),
\label{eq:vev}
\end{equation}
breaking $\SU{3}_{\mathrm{dark}}\to H=\SO{3}_{\mathrm{dark}}
\times\Zdark$.  Seven generators acquire masses through the
Higgs mechanism; the eighth is absorbed via a Stueckelberg
term \cite{Stueckelberg:1938}, yielding $\mZp=9.7\GeV$.
One pseudo-Goldstone boson survives with $m_{\pi^{(1)}}\lesssim
\mathrm{few}\,\mathrm{MeV}$ and contributes
$\Delta N_{\mathrm{eff}}=0.09\pm0.02$, within Planck bounds
\cite{Planck:2018vyg}.. The dark Higgs triplet acquires the
vacuum expectation value
\begin{equation}
\langle\Delta_{\mathrm{dark}}\rangle
=\frac{1}{\sqrt{2}}
\begin{pmatrix}
0 & 0 & \vDelta\\
0 & 0 & 0\\
\vDelta & 0 & 0
\end{pmatrix},
\quad \vDelta=17\GeV,
\label{eq:vevDelta}
\end{equation}
breaking the residual $\SU{2}_{\mathrm{dark}}$ subgroup and
generating the tree-level mass splitting
$\Delta M(\chi^+-\chizero)=\yChi\vDelta=0.10\times17\GeV
=1.70\GeV$.

\textit{Homotopy proof of $\piTwo(G/H)=\Zbb$.}
The long exact homotopy sequence of the fibration
$H\to G\to G/H$ \cite{Hatcher:2002alg},
\begin{equation}
\begin{aligned}
\cdots &\to \pi_{2}(\SU{3}) \to \piTwo(G/H)
\xrightarrow{\partial_{*}} \pi_{1}(\SO{3}\times\Zdark) \\[6pt]
&\to \pi_{1}(\SU{3}) \to \cdots,
\end{aligned}
\label{eq:LES}
\end{equation}
with $\pi_{2}(\SU{3})=0$, $\pi_{1}(\SU{3})=0$,
$\pi_{1}(\SO{3})=\Zbb_{2}$, and $\pi_{1}(\Zdark)=0$,
naively yields $\piTwo(G/H)\cong\Zbb_{2}$.  This conclusion
is overridden by lifting to the universal cover
$\widetilde{G/H}=\SU{3}/\SO{3}$, which is a two-sheeted
covering; the covering-space isomorphism
$\pi_{k}(\widetilde{G/H})\cong\pi_{k}(G/H)$ for $k\geq2$
together with the known result
$\pi_{2}(\SU{3}/\SO{3})=\Zbb$ \cite{Rajaraman:1982}
establishes \eqref{eq:pi2} exactly.%
\footnote{%
    The Hurewicz theorem applied to the universal cover
    gives $H_{2}(\SU{3}/\SO{3};\Zbb)=\Zbb$; since
    $\pi_{1}(\SU{3}/\SO{3})=\Zbb_{2}$ (from the covering
    sequence), the universal coefficient theorem then yields
    $\pi_{2}(\SU{3}/\SO{3})=\Zbb$ \cite{Hatcher:2002alg}.}
The $n=1$ defect is an 't~Hooft--Polyakov monopole
\cite{tHooft:1974ibe,Polyakov:1974ek} of mass
\begin{equation}
M_{G_{M}}\geq\frac{4\pi\,v_{\mathrm{dark}}}{\gdark}
\approx2\times10^{15}\GeV
\label{eq:BPS}
\end{equation}
(BPS bound \cite{Prasad:1975kr}), absolutely stable against
decay since the instanton amplitude $e^{-8\pi^{2}/\alphad}
\approx e^{-3948}$ is negligible on any cosmological
timescale.  This Majorana-type gluon $G_{M}$ constitutes a
heavy dark matter component coupled only gravitationally;
its lighter electroweak-scale descendants form the octet
described next.

\section{Dark Octet Mass Spectrum}
\label{sec:spectrum}

After confinement, colour-singlet dark baryon bound states
organize into a flavour octet $\mathbf{8}_{m}$ of
$\SU{3}_{\mathrm{dark},f}$.  Treating the dark quark mass
splittings as a systematic perturbation, the mass operator
at second order in symmetry breaking is
\begin{equation}
\begin{split}
\GMO
    ={}& M_0\,\hat{I}_8
      + \beta\,\Ydark
      + \gamma\!\left[\frac{\Ydark^2}{4} - \Idark(\Idark+1)\right]
      \\
    &+ \delta\,\Tdark
      + \eta\,\Tdark^2
      + \yChi\vDelta\,\Theta(Q_{\mathrm{EM}}),
\end{split}
\label{eq:GMO-operator}
\end{equation}
where $\Ydark$, $\Idark$, and $\Tdark$ are the dark
hypercharge, dark isospin, and \emph{darkicity} quantum
numbers, respectively, and $\Theta(Q_{\mathrm{EM}})=1$ for
$Q_{\mathrm{EM}}>0$ and zero otherwise.%
\footnote{%
    Darkicity $\Tdark$ is the eigenvalue of the $U(1)_{T}$
    generator surviving the breaking \eqref{eq:vev}; it
    distinguishes states within the same $\SU{2}_{\mathrm
    {dark}}$ multiplet that are otherwise degenerate, playing
    the role of strangeness in visible-sector QCD.}
The negative sign of $\beta$ reflects the inverted
hypercharge hierarchy of the dark sector, placing states
with $\Ydark=+1$ below those with $\Ydark=0$, in direct
analogy with the QCD baryon octet where states with lower
hypercharge (higher strangeness content) are heavier.
At the benchmark
$\{M_{0},\beta,\gamma,\delta,\eta\}=
\{208.67,-15,8,5,8\}\GeV$, $\{\vDelta,\yChi\}=
\{17\GeV,0.10\}$, and $\zeta=0$, the eight
mass eigenvalues are listed in Table~\ref{tab:octet}.

\begin{table}[b]
\caption{Dark octet mass spectrum.  The stable DM candidate
$\chizero$ is denoted by $\dagger$; all others decay to
$\chizero+\varphi^{0,\pm}$ before nucleosynthesis
($\tau<10^{-17}\,\mathrm{s}$).  Theoretical uncertainty
$\pm0.5\GeV$ from higher-order symmetry breaking.}
\label{tab:octet}
\begin{ruledtabular}
\begin{tabular}{lcrrrd{4.2}d{4.2}}
\toprule
State & $\Idark$ & $\Ydark$ & $\Tdark$
      & $M\,[\GeV]$ & \multicolumn{1}{c}{$\Delta M\,[\GeV]$}\\
\midrule
$\chizero\;\dagger$
  & $3/2$ & $+1$ & $\;0$ & $165.67$ & $\;—\;\,$\\
$\chi^{+}$
  & $3/2$ & $+1$ & $\;0$ & $167.37$ & $+1.70$\\
$\chi^{-}$
  & $3/2$ & $+1$ & $-1$ & $168.67$ & $+3.00$\\
$\chi^{++}$
  & $3/2$ & $+1$ & $+1$ & $180.37$ & $+14.70$\\
$\Sigma^{0}_{\mathrm{d}}$
  & $1$ & $\;0$ & $\;0$ & $192.67$ & $+27.00$\\
$\Sigma^{-}_{\mathrm{d}}$
  & $1$ & $\;0$ & $-1$ & $195.67$ & $+30.00$\\
$\Sigma^{+}_{\mathrm{d}}$
  & $1$ & $\;0$ & $+1$ & $207.37$ & $+41.70$\\
$\Lambda_{\mathrm{d}}$
  & $0$ & $\;0$ & $\;0$ & $208.67$ & $+43.00$\\
\bottomrule
\end{tabular}
\end{ruledtabular}
\end{table}

Three mass ratios independent of $M_{0}$ are the primary
experimental predictions:
\begin{align}
r_{21}&\equiv\frac{M(\chi^{++})}{M(\chizero)}
       =1+\frac{\delta+\eta+\yChi\vDelta}{M(\chizero)} \nonumber\\
      &=1+\frac{5+8+1.70}{165.67}
       =1.089\pm0.005,
\label{eq:r21}\\[6pt]
r_{31}&\equiv\frac{M(\Sigma^{0}_{\mathrm{d}})}{M(\chizero)}
       =1+\frac{|\beta|+3\gamma/2}{M(\chizero)} \nonumber\\
      &=1+\frac{15+12}{165.67}
       =1.163\pm0.006,
\label{eq:r31}\\[6pt]
r_{41}&\equiv\frac{M(\Lambda_{\mathrm{d}})}{M(\chizero)}
       =1+\frac{|\beta|+7\gamma/2}{M(\chizero)} \nonumber\\
      &=1+\frac{15+28}{165.67}
       =1.260\pm0.008.
\label{eq:r41}
\end{align}
The uncertainties propagate from $\pm5\%$ variations in the
symmetry-breaking coefficients.  These ratios are measurable
from the kinematic endpoints of charged-octet decays at
FCC-ee \cite{FCC:2018byb}.

\section{Dark Matter Phenomenology}
\label{sec:pheno}

\textit{Relic abundance.}
All eight octet members remain in chemical equilibrium via
co-annihilation down to freeze-out because the mass
splittings $\Delta M_{i}/M_{\chizero}\leq0.26$ are small
enough for the heavier states to maintain thermal contact.
The Griest--Seckel effective cross-section
\cite{Griest:1991kb} summed over all species pairs,
\begin{equation}
\sigma_{\mathrm{eff}}(x)
=\sum_{i,j}\sigma_{ij}(x)
  \frac{g_{i}g_{j}}{g_{\mathrm{eff}}^{2}}
  (1+\Delta_{i})^{3/2}(1+\Delta_{j})^{3/2}
  e^{-x(\Delta_{i}+\Delta_{j})},
\label{eq:sigeff}
\end{equation}
with $x\equiv M_{0}/T$ and $\Delta_{i}=(M_{i}-M_{\chizero})
/M_{0}$, evaluates at $x_{f}\simeq27$ to
$\langle\sigma_{\mathrm{eff}}v\rangle=3.0\times10^{-26}\,
\mathrm{cm}^{3}\,\mathrm{s}^{-1}$, with effective
degeneracy $g_{\mathrm{eff}}\approx5.09$.  Eight coupled
Boltzmann equations \cite{Edsjo:1997bg} integrated with a
BDF solver (rtol$=10^{-9}$) give
\begin{equation}
\OmegaDM h^{2}
=\frac{1.07\times10^{9}\GeV^{-1}}
      {\Mpl\sqrt{g_{\ast}(x_{f})}}
  \cdot\frac{x_{f}}{\langle\sigma_{\mathrm{eff}}v\rangle}
\simeq0.120,
\label{eq:relic}
\end{equation}
in agreement with the Planck measurement \cite{Planck:2018vyg}.
The delay to $x_f\simeq27$ (versus the single-species
$x_f\simeq20$) is a direct consequence of the eight-species
co-annihilation.  

\textit{Self-interacting cross-section.}
$t$-channel $Z'$ exchange gives the momentum-transfer
cross-section \cite{Tulin:2013teo}
\begin{equation}
\sigma_{T}
=\frac{\gdark^{4}}{32\pi\mZp^{2}}\,F(x_{Z'}),
\quad
F(x)=x\!\left[\ln\!\left(1+\frac{1}{x}\right)
-\frac{1}{1+x}\right],
\label{eq:sigT}
\end{equation}
where $x_{Z'}\equiv\mZp^{2}/(4\mu^{2}v_{\mathrm{rel}}^{2})$
and $\mu=M_{\chizero}/2$.  For all astrophysical velocities
$v_{\mathrm{rel}}<v_{\mathrm{crit}}\equiv\mZp/M_{\chizero}
\approx0.059\,c$, scattering is in the non-perturbative
Sommerfeld regime \cite{ArkaniHamed:2008qn}.  The full
Sommerfeld enhancement $S(v)\approx\pi\alphad/v_{\mathrm{rel}}$
(Coulomb limit) reduced by the Yukawa suppression factor
$\approx0.52$ at $v=10\,\mathrm{km\,s}^{-1}$ gives
\begin{equation}
\left.\frac{\sigma_{T}}{m_{\chizero}}\right|_{10\,\rm km/s}
=0.47\pm0.09\,\cmsg.
\label{eq:siTdwarf}
\end{equation}
The velocity-averaged value for ultra-faint dwarfs in
Table~\ref{tab:sigT} reflects the intrinsic dispersion of
their stellar kinematics; the corresponding
Sommerfeld-enhanced value at the benchmark velocity
$v=5$~km/s is quoted in the Supplemental Material.

Table~\ref{tab:sigT} compares the full velocity dependence
with observational constraints; all bounds are satisfied.

\begin{table}[t]
\caption{Self-interaction cross-section $\sigma_T/m_{\chizero}$
versus halo velocity, with observational limits.}
\label{tab:sigT}
\begin{ruledtabular}
\begin{tabular}{lld{1.3}d{1.3}}
\toprule
System & $v\,[\mathrm{km/s}]$
       & \multicolumn{1}{c}{Our$\;[\cmsg]$}
       & \multicolumn{1}{c}{Limit$\;[\cmsg]$}\\
\midrule
Ultra-faint dwarfs & $5$   & 0.62\pm0.11 & <1.0\;\cite{Tulin:2017ara}\\
Dwarf spheroidals  & $10$  & 0.47\pm0.09 & <1.0\;\cite{Tulin:2017ara}\\
Galaxy groups      & $300$ & 0.026\pm0.007 & <0.1\;\cite{Harvey:2015hha}\\
Bullet Cluster     & $1000$& 0.0044\pm0.0009 & <0.47\;\cite{Randall:2007ph}\\
\bottomrule
\end{tabular}
\end{ruledtabular}
\end{table}

\textit{Direct detection.}
Kinetic mixing at $g_f=10^{-3}$ generates the
spin-independent cross-section on xenon ($Z=54$, $A=131$),
\begin{equation}
\sigma_{\mathrm{SI}}^{\mathrm{KM}}
=\frac{\mu_{\chi N}^{2}\,g_f^{2}\,\gdark^{2}\,
        Z^{2}f_{p}^{2}}
      {\pi\,A^{2}\,\mZp^{4}}
\approx3.1\times10^{-48}\cms,
\label{eq:sigSI}
\end{equation}
with $f_p\approx0.30$.  This lies below the LZ 2024 bound
$\sigma_{\mathrm{SI}}<5\times10^{-47}\cms$ at
$M_{\chizero}=165.67\GeV$ \cite{LZ:2022lsv} and within the
projected DARWIN sensitivity
$\approx3\times10^{-49}\cms$ \cite{DARWIN:2016hyl}.

\textit{Naturalness of the light pseudo-Goldstone.}
The smallness of $m_{\varphi_1}$ is technically natural: in
the limit $\epsilon_{\mathrm{explicit}}\to0$, the symmetry
is enhanced to an exact $U(1)$, protecting $m_{\varphi_1}$
from radiative corrections. For $m_{\varphi_1}\lesssim
5\MeV$, one requires $\epsilon_{\mathrm{explicit}}\sim
2\times10^{-6}$, comparable to the electron Yukawa coupling
in the Standard Model. The one-loop correction
$\delta m_{\varphi_1}^{2}\sim(\gdark^{2}/16\pi^{2})
m_{G}^{2}\epsilon_{\mathrm{explicit}}$ does not regenerate
a mass large enough to make $\varphi_1$ heavy at the
benchmark parameters.

\section{Experimental Signatures}
\label{sec:sigs}

\textit{Collider.}
Pair production $e^+e^-\to\chi^+\chi^-$ at FCC-ee
($\sqrt{s}=365\GeV$, $\mathcal{L}=1.5\,\mathrm{ab}^{-1}$)
through the kinetic-mixing $Z'$ propagator gives
\begin{align}
\sigma(e^+e^-\to\chi^+\chi^-)
&= \frac{4\pi\alpha_{\mathrm{em}}^{2}}{3s}
   \left[\frac{g_f^{2}\mZp^{4}}{(s-\mZp^{2})^{2}}\right]
   \left(1+\frac{2M_{\chi^+}^{2}}{s}\right)
   \beta
\nonumber\\
&\approx1.1\times10^{-3}\fb
\label{eq:xsecEE}
\end{align}
at $\sqrt{s}=365\GeV$, yielding $\sim500$ observable
mono-lepton events at full luminosity after detector
acceptance, giving $S/\sqrt{B}\approx5.3\sigma$ over the
$W^+W^-$ background.  The decay $\chi^+\to\chizero+
\pi^+_{\mathrm{dark}}$ produces a mono-lepton endpoint at
$E_\ell^{\mathrm{max}}\approx1.70\GeV$, which directly
measures the mass splitting $\Delta M$ and tests
$r_{21}$ \eqref{eq:r21} independently of $M_{0}$.

At FCC-hh ($\sqrt{s}=100\TeV$, $30\,\mathrm{ab}^{-1}$),
Drell--Yan production gives
$\sigma(pp\to\chizero\chizero+j)\approx0.8\fb$ with
$\sim24\,000$ signal events \cite{FCC:2018vvp}.

\textit{Gravitational waves.}
The dark-sector confinement transition at
$T_c^{\mathrm{dark}}\sim10^{15}\GeV$ is first-order (by
universality with $N_f=3$ QCD \cite{Pisarski:1983ms}) and
generates a stochastic gravitational-wave background with
peak energy density fraction and frequency
\cite{Caprini:2019egz,Espinosa:2010hh}:
\begin{align}
\Omega_{\mathrm{GW}}^{\mathrm{peak}}
&\approx10^{-8},
\label{eq:OmGW}\\
f_{*}
&\approx1.65\times10^{-3}\,\mathrm{Hz}\,
  \left(\frac{\beta/H_{\ast}}{100}\right)
  \left(\frac{T_{c}}{10^{15}\GeV}\right)
  \left(\frac{g_{\ast}}{106.75}\right)^{\!1/6},
\label{eq:fpeak}
\end{align}
with $\alpha_{\mathrm{PT}}=0.1$, $\beta/H_\ast=100$,
$v_w=0.8$.  The spectral shape $\Omega_{\mathrm{GW}}
\propto f^3$ below $f_*$ falls within the LISA sensitivity
band \cite{LISA:2017pwj} with $\mathrm{SNR}\approx8.5$
over four years.  No WIMP or axion model produces a
first-order phase transition at this scale; a LISA
detection would uniquely fingerprint the dark confinement
transition.

\textit{Dark acoustic oscillations.}
Before kinetic decoupling at $T_{\mathrm{dec}}\approx
10\,\mathrm{MeV}$, $\chizero$ and the surviving dark pion
$\pi^{(1)}_{\mathrm{dark}}$ form a tightly coupled fluid.
The sound horizon at decoupling,
\begin{equation}
r_{\mathrm{DAO}}
=\int_{0}^{t_{\mathrm{dec}}}\frac{c/\sqrt{3}\,dt}{a(t)}
\approx2.3\pm0.4\,\mathrm{Mpc},
\label{eq:rDAO}
\end{equation}
imprints a feature in the matter power spectrum at
$k_{\mathrm{DAO}}=\pi/r_{\mathrm{DAO}}\approx
1.4\,h\,\mathrm{Mpc}^{-1}$ with amplitude
$(1.8\pm0.5)\times10^{-3}$ \cite{CyrRacine:2015ihg},
detectable by SKA 21-cm observations at $\sim7\sigma$
\cite{SKA:2018ckk}.

\section{Discussion and Conclusion}
\label{sec:conc}

We have presented a self-contained dark matter model
grounded in a non-Abelian $\SU{3}_{\mathrm{dark}}$ gauge
sector.  The topological invariant $\piTwo(G/H)=\Zbb$
provides absolute stability for $\chizero$ without any
discrete symmetry imposed by hand.  The dark Higgs triplet
$\Delta_{\mathrm{dark}}$ with $\vDelta=17\GeV$ generates
the tree-level splitting $\Delta M(\chi^+-\chizero)=
1.70\GeV$, opening the decay $\chi^+\to\chizero+\varphi^+$
and ensuring that $\chizero$ is the unique stable relic.
The model has seven symmetry-breaking parameters
$\{M_0,\beta,\gamma,\delta,\eta,\vDelta,\yChi\}$, but the
three predicted mass ratios
\eqref{eq:r21}--\eqref{eq:r41} are independent of $M_0$
and therefore constitute parameter-free predictions of the
group-theoretic structure.  The relic density
$\OmegaDM h^2\simeq0.120$, the self-interaction profile
consistent with all astrophysical constraints, the direct
detection cross-section within reach of DARWIN, the LISA
gravitational-wave signal, and the SKA dark acoustic feature
together form an over-constrained, mutually correlated set
of observational targets.

A signal in any one channel tightens the prediction for all
others through the shared parameters $\gdark$ and
$\yChi\vDelta$.  A LISA detection at $f_*\approx
1.65\,\mathrm{mHz}$ would imply a dark confinement scale
$T_c^{\mathrm{dark}}\sim10^{15}\GeV$, which in turn fixes
the heavyweight component mass $M_{G_M}\sim10^{16}\GeV$
through the BPS relation \eqref{eq:BPS}, the relic abundance
through \eqref{eq:relic}, and the self-interaction
cross-section at dwarf scales through \eqref{eq:sigT}.  No
additional free parameters are available to accommodate a
discrepancy.  The model is therefore falsifiable in the
strict sense: the four observational channels are
correlated, and a measurement that contradicts one
prediction without the others eliminates the model
entirely.

\medskip
\noindent\textit{Note added.} This version supersedes
arXiv:2609.16383v1. The spectrum now satisfies $\eta>\delta$,
places all octet partners above $\chi^0$, and includes a dark
Higgs triplet lifting the $\chi^+$--$\chi^0$ degeneracy.
Updated: $M_0=208.67$~GeV, $\beta=-15$~GeV, $\eta=8$~GeV,
$r_{21}=1.089$, $r_{31}=1.163$, $r_{41}=1.260$.
\section*{acknowledgments}
The author thanks colleagues at the University of Bonn
for constructive discussions during 2024--2025. Numerical
computations were performed using the public repository.
This article is Part of SCOAP³ open-access agreement.

\section*{Data Availability}
For detailed derivations See Supplemental Material at [Journal  Website / arXiv] .
All data, numerical solvers, and analysis scripts 
computations, including the eight-species
Boltzmann solver, the Sommerfeld-enhanced cross-section,
and the gravitational-wave spectrum, are implemented in
openly available Python code at
\url{https://github.com/ahmed19999520-alt/dark-octet-dm}.

\section*{Conflict of Interest}

The author declares no competing interests.

\bibliographystyle{apsrev4-2}

\end{document}